\documentclass[%
 amsmath,amssymb,
 aps, physrev,
]{revtex4-2}

\usepackage{graphicx}
\usepackage{dcolumn}
\usepackage{bm}
\usepackage[mathlines]{lineno}
\usepackage{xcolor}

\begin{document}

\preprint{APS/123-QED}


\title{\textbf{Coherent Structure Transport in Turbulent Axisymmetric Pipe Expansions}}

\author{Jibu Tom Jose}
 \author{Gal Friedmann}
 \author{Omri Ram}
 \email{omri.ram@technion.ac.il}
\affiliation{ 
Faculty of Mechanical Engineering, Technion - Israel Institute of Technology Haifa Israel
}%


\begin{abstract}

Turbulent separated flows in axisymmetric expansions can sustain fundamentally different transport organization despite nearly identical mean topology. Using stereo-PIV and time-resolved planar PIV, we compare abrupt $90^\circ$ (step) and gradual $45^\circ$ (wedge) axisymmetric expansions at step height Reynolds numbers of 25000 and 35000. Despite similar reattachment lengths, near-separation turbulence differs, with the wedge exhibiting higher turbulent kinetic energy over a broader shear layer, while the step confines production to a thinner region near the corner, where a secondary vortex weakens momentum and fluctuations. The spatial spectra reveal a pronounced spectral hump in the out-of-plane velocity fluctuations near separation. This feature is consistently observed across all cases and reflects the expansion effects on the redistribution of fluctuation energy associated with the interaction between the separating shear layer and the recirculating flow. Temporal spectra show no geometry-specific dominant frequencies, and space-time correlations indicate similar normalized convection velocities across all cases. The primary effect of geometry, therefore, does not lie in the characteristic scales or transport speeds, but in the spatial organization and persistence of coherence. The step cases exhibit stronger spectral concentration, longer local integral time scales, and a broader distribution of space-time correlations in convection velocities associated with momentum-depleted return flow. Finite-time Lyapunov exponent (FTLE) fields confirm that these differences extend to material transport, as the wedge produces larger and less fragmented deformation regions, while the step yields a more segmented pattern that persists downstream of reattachment.
\end{abstract}

\maketitle

\section{Introduction}
\label{sec:intro}

Sudden expansions in internal flows produce separation at the expansion corner and generate a free shear layer that grows downstream, enclosing a recirculation zone before reattaching to the wall. Experimental investigations of pipe expansions have characterized the structure of the separated region and its downstream recovery primarily through pressure measurements and laser doppler velocimetry \citep{back1972shear, morrison1988three, durrett1988radial, stieglmeier1989experimental}, while subsequent measurements have examined the fundamental development of the separated shear layer in axisymmetric geometries \citep{gould1990investigation, furuichi2003spatial, hammad1999piv, goharzadeh2009experimental}. Planar back-facing step (BFS) experiments, which closely mirror the flow topology of axisymmetric expansions, have established that reattaching shear layers are structured by organized vortical motions whose roll-up and interaction regulate entrainment and recovery \citep{browand1966experimental, winant1974vortex, troutt1984organized, eaton1982low}. Coherent structures generated near separation have been shown to convect downstream and modulate the instantaneous reattachment process \citep{cantwell1983experimental, devenport1993experimental, jovic1996experimental, hudy2007stochastic}. Correlation-based analyses using two-point velocity correlations, conditional averaging, and linear stochastic estimation have shown that a limited number of spatially coherent motions dominate momentum transport and fluctuation dynamics in separated flows \citep{kasagi1995three, cole1998applications, lee2002multiple}. Complementary numerical studies using direct numerical simulation and large eddy simulation have further clarified the underlying dynamics, revealing the coexistence of shear layer instabilities, vortex interactions, and broader unsteady motions in separated configurations \citep{le1997direct, schafer2009dynamics, pont2019direct, nguyen2019perturbation, xiang2024drldl}.

Axisymmetric expansions introduce additional complexity due to curvature and out-of-plane coherence that can modify the global organization. Early measurements in circular pipe expansions have documented three-dimensional effects and radial redistribution of momentum \citep{durrett1988radial, stieglmeier1989experimental,gould1990investigation, devenport1993experimental}. Subsequent spatially resolved particle image velocimetry (PIV) measurements, including refractive-index-matched configurations, have enabled detailed access to velocity-field structure in axisymmetric geometries \citep{hammad1999piv, mak2007near, amini2012investigation, scharnowski2015investigation,jose2026effect}. These studies have provided a wealth of information on the mean structure and turbulence statistics of separated pipe flows, yet few investigations have examined how coherent motions are organized simultaneously in space and time in turbulent axisymmetric expansions.

The expansion angle has been shown to influence the separation topology and the downstream shear-layer development. Numerical studies have demonstrated that gradual expansions redistribute turbulence production and modify shear layer growth relative to abrupt steps \citep{teyssandiert1974analysis, selvam2015localised, ahmadpour2016numerical, danane2020effect}, while also altering the associated loss mechanisms \citep{choi2016numerical, lebon2018new, zhou2025large, rosa2025pressureloss}. However, how the expansion geometry reorganizes the dynamics of the separated shear layer, beyond its effect on mean turbulence production, remains poorly understood.
 
In our recent study~\citep{jose2026effect}, we established through high-resolution refractive-index-matched stereo-PIV that abrupt($90^\circ$) and gradual ($45^\circ$) axisymmetric expansions at matched Reynolds numbers produce fundamentally different spatial distributions of turbulence production and Reynolds-stress anisotropy near separation. The governing mechanism is the geometry-induced modulation of the return flow; in the gradual expansion the return flow follows the sloped wall closely and impinges obliquely on the incoming free-stream, generating a distributed region of strong shear and elevated turbulence production, whereas in the abrupt expansion a secondary recirculation vortex at the expansion foot depletes return-flow momentum and the subsequent $90^\circ$ redirection further suppresses its interaction with the free-stream, confining production to a thin, intense band near the separation corner. These results identify where fluctuation energy is injected into the flow, but, because they were based on solely on statistically converged mean fields, they could not resolve how that energy is subsequently organized in space and time, how coherent motions structure transport downstream, or whether the geometry-dependent differences established near separation persist through reattachment as material transport signatures. The present study addresses these open questions directly. 

This study investigates the spatial, temporal, and space-time organization of turbulence in turbulent axisymmetric expansions using high-resolution stereo PIV measurements to quantify turbulence fields and spatial coherence, together with high-speed planar PIV to resolve the temporal dynamics of the separated shear layer. Two-point correlations and spectra are used to extract coherence scales and convection characteristics, following established methodologies in turbulent structure analysis \citep{tennekes1972first, Pope_2000,mckeon2004friction, marusic2010wall}. In addition, Lagrangian coherent structure (LCS) diagnostics provide a transport-based perspective by identifying material pathways that organize mixing in separated flows \citep{shadden2005definition, mathur2007uncovering, haller2015lagrangian, huang2022lagrangian}. By directly comparing abrupt and gradual axisymmetric expansions at matched Reynolds numbers, we determine that the expansion geometry fundamentally alters the dynamical content of the separated shear layer and primarily redistributes the organization and persistence of coherent motions formed at the separation. Through this transport-oriented framework, the present work extends beyond mean topology and turbulence production to quantify how geometry reorganizes coherent transport pathways in axisymmetric separated flows.
\section{Experimental methodology}
\subsection{Facility and test configurations}

\begin{figure*}
    \centering
    \includegraphics[width=\linewidth]{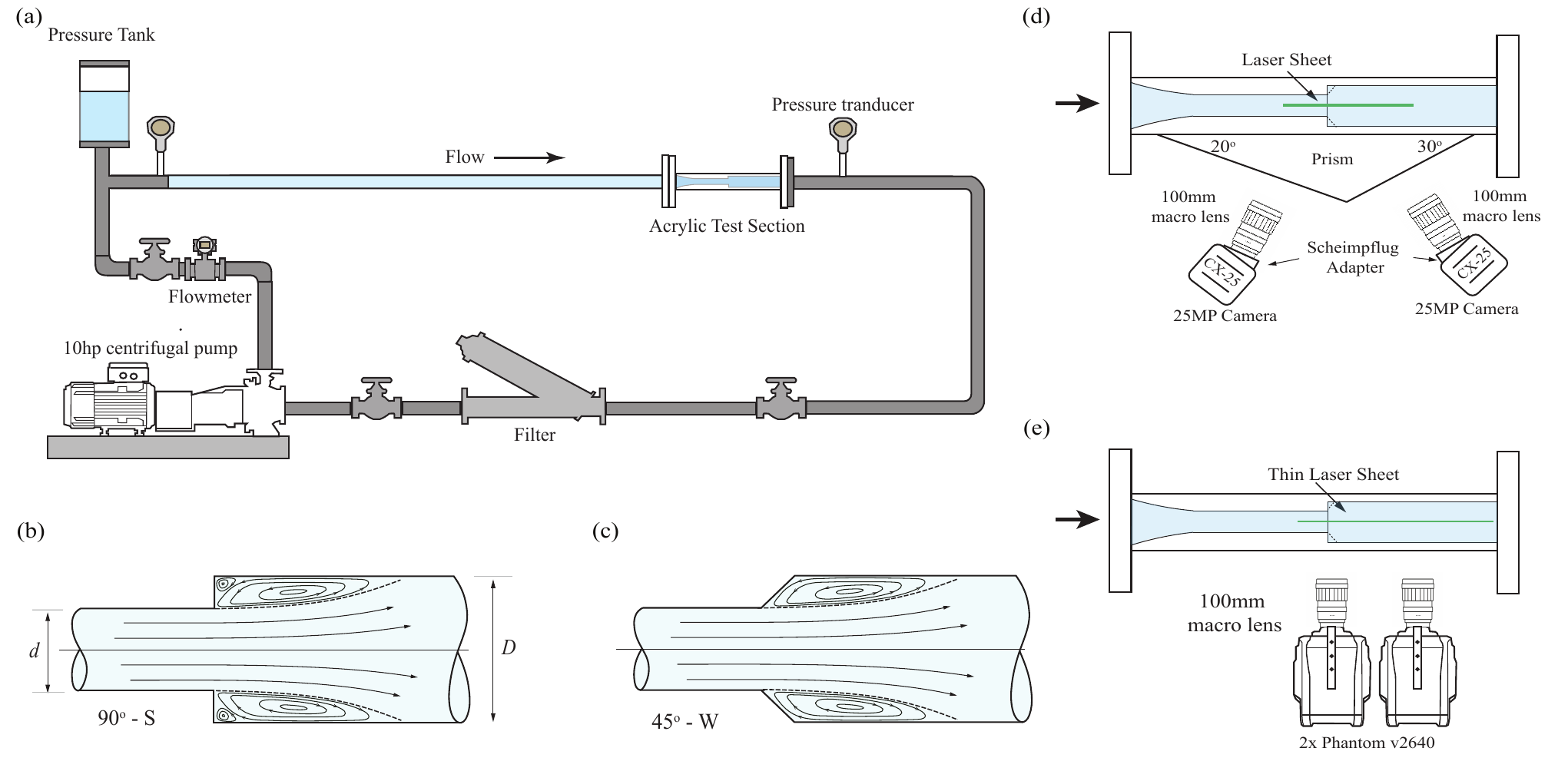}
    \caption{(a) Schematic of the refractive-index-matched water tunnel facility. A schematic of the time-averaged flow field in a pipe with axisymmetric area expansion through a (b) $90^\circ$ step (S) and (c) $45^\circ$ wedge (W) expansion. (d) stereo-PIV configuration using two high-resolution cameras; (e) side-by-side high-speed planar PIV configuration used to extend the streamwise measurement domain.}
    \label{fig:channel_PIV}
\end{figure*}

Experiments are conducted in the refractive-index-matched water tunnel facility at the Transient Fluid Mechanics Laboratory at Technion. A schematic overview of the facility and the PIV camera configurations is shown in Fig.~\ref{fig:channel_PIV}(a). The closed-loop water tunnel is driven by a 10~hp centrifugal pump, and provides volumetric flow rates up to 130~m$^3$/h. The flow rate is regulated using a variable-frequency drive, enabling steady operation over the freestream velocity range considered in this study. To ensure optical stability and minimize mechanical vibrations, the maximum flow velocity during experiments is limited to 6~m/s. An electromagnetic flowmeter with an accuracy of $\pm0.5\%$ is used to monitor the volumetric flow rate. System pressure is controlled via a transparent pressurization tank mounted at the top of the loop, which serves as a pressure-stabilizing volume and safety buffer against over-pressurization. Nitrogen pressurizes the gas phase to limit oxidation of the working fluid. Prior to measurements, the loop is de-aerated using a vacuum pump to suppress bubble formation at elevated flow rates. Two pressure transducers continuously monitor the operating pressure to ensure safe and stable operation of the water tunnel. 

Upstream of the test section, the flow passes through a 2000~mm-long straight acrylic pipe of diameter 40~mm, corresponding to more than 50 pipe diameters of development length, ensuring a fully developed turbulent inlet condition at the entrance to the expansion. The test section consists of a 300~mm-long interchangeable acrylic test section, allowing different internal geometries to be examined under identical upstream conditions. A 62–63\% aqueous solution of sodium iodide (NaI) is used as the working fluid, as its refractive index matches that of the acrylic window, enabling unobstructed optical access across curved surfaces. At $20^\circ$C, the NaI has a kinematic viscosity of $1.1\times10^{-6}$~m$^2$/s and a specific gravity of 1.82 \citep{bai2014refractive}. The external surfaces of the test sections are machined flat, and acrylic prisms maintain the camera optical axes at normal optical incidence during stereo-PIV experiments, thereby minimizing distortions. Two axisymmetric expansion geometries are investigated, as shown in Fig.~\ref{fig:channel_PIV}(b) and (c). The first configuration consists of an abrupt $90^\circ$ expansion from an inlet diameter of 25~mm to a downstream diameter of 40~mm, corresponding to a step height of $h=7.5$~mm and an area expansion ratio of 2.56. The second configuration has the same area ratio, but uses a gradual $45^\circ$ expansion. For both geometries, experiments are conducted at step-height Reynolds numbers $Re_h = U_m h / \nu$ of 25000 and 35000, corresponding to inlet bulk velocities of 3.6~m/s and 5.1~m/s, respectively. The four cases are denoted as S-25, S-35 ($90^\circ$ step expansion) and W-25, W-35 ($45^\circ$ wedge expansion).

Velocity measurements downstream of the expansion are obtained using a combination of stereo-PIV and time-resolved planar PIV. The stereo-PIV measurements provide statistically converged three-component velocity fields for characterizing the mean flow and turbulence statistics. In contrast, the time-resolved planar PIV measurements resolve the temporal evolution of the separated flow and coherent structures, enabling analysis of spatio-temporal organization and transport beyond mean quantities. For the stereo-PIV measurements, two high-resolution cameras (LaVision Imager CX-25, 25~MP) equipped with 100~mm macro lenses are arranged in a stereoscopic configuration, as shown in Fig.~\ref{fig:channel_PIV}(d). The cameras have a field of view of approximately $107\times54$~mm$^2$ at a magnification of 55.4~pixels/mm. Illumination is provided by a dual-cavity Nd:YAG laser (Quantel EverGreen 532-200), forming a light sheet approximately 1~mm thick. Each case is recorded for 200~s at an acquisition rate of 15~Hz, yielding 3000 statistically independent realizations for ensemble averaging. Time-resolved planar PIV measurements are performed using two synchronized high-speed cameras (Phantom v2640, 4~MP) arranged side-by-side to extend the streamwise measurement domain as shown in Fig.~\ref{fig:channel_PIV}(e). This configuration provides a combined field of view of approximately 148~mm in the streamwise direction, corresponding to about 16.5 step heights downstream of the expansion. The cameras operate at 8000~Hz with a spatial resolution of $3665\times1348$~pixels and a magnification of 24.3~pixels/mm. A dual-head Nd:YAG laser (Photonics Industries DM532-100) is used as the illumination source, producing a light sheet approximately 0.3~mm thick. Each high-speed acquisition consists of approximately 18700 images, corresponding to a recording length of 2.3~s. In both measurement configurations, silver-coated hollow glass spheres with a mean diameter of 13~$\mu$m are used as tracer particles, resulting in a particle Stokes number of approximately $8\times10^{-4}$. A summary of the PIV imaging parameters is provided in Table~\ref{table:PIV}.

\begin{table}
    \centering
    \begin{tabular}{lcc}
        \hline
        \textbf{Parameter} &
        \textbf{Stereo PIV} &
        \textbf{Planar high-speed PIV} \\
        \hline
        Field of view (mm) & $107\times54$ & $148\times42$ \\
        Image resolution (px) &  $5928\times2992$   & $3665\times1348$ \\ 
        Acquisition rate (Hz) & 15 & 8000 \\
        Vector spacing ($\mu$m) & 217 & 494 \\
        Number of snapshots & 3000 & 18700 \\
        Measurement duration (s) & 200 & 2.3 \\
        Cameras & 2$\times$ CX-25 & 2$\times$ v2640 \\
        \hline
    \end{tabular}
    \caption{
    Summary of imaging and acquisition parameters for the stereo-PIV and time-resolved planar PIV measurements.
    }
    \label{table:PIV}
\end{table}

\subsection{Data processing and analysis}
\label{ssec:analysis}

Velocity fields are computed using standard cross-correlation PIV using LaVision DaVis\texttrademark~11. Raw images are pre-processed using sliding background subtraction to suppress stationary reflections and enhance particle contrast. Stereo-PIV calibration is performed using a multi-plane target, followed by a self-calibration to refine the mapping functions and reduce residual disparity errors. For the side-by-side planar PIV configuration, a custom two-dimensional calibration plate extending across the full field of view is used, ensuring sufficient overlap between the two cameras. For both stereo-PIV and planar PIV datasets, the final interrogation window size is $24\times24$~pixels with 50\% overlap. For the stereo-PIV measurements, ensemble-averaged flow fields are also computed using a sum-of-correlation approach with $12\times12$ windows and 75\% overlap over the full set of 3000 realizations. This procedure improves the spatial resolution of the mean flow without altering instantaneous turbulence statistics. The resulting vector spacing is 217~$\mu$m for standard stereo-PIV fields and 54~$\mu$m for sum-of-correlation fields. The time-resolved planar PIV data yield a vector spacing of 494~$\mu$m and a temporal resolution of 125~$\mu$s. For the present data set, the Kolmogorov length scale $\eta$ is approximately 43~$\mu$m, yielding a spatial resolution of $\sim 5\eta$ for stereo-PIV data and $\sim 11.5\eta$ for the time-resolved data. The corresponding Kolmogorov time scale $\tau_\eta$ is approximately 1.7~ms, and the temporal resolution of the high-speed data is $\sim 0.07\tau_\eta$.

Owing to the axisymmetric nature of the test section, a cylindrical coordinate system is employed throughout the paper, with the streamwise direction denoted by $z$ and the radial direction by $r$, measured from the pipe centerline. The location $z=0$ corresponds to the expansion corner. Spatial coordinates are normalized by the step height $h$, denoted as $r^*=r/h$ and $z^*=z/h$. Under this normalization, the step corner is located at $r^*=1.67$, and the wall of the expanded section lies at $r^*=2.67$ downstream of the expansion. Instantaneous velocity fluctuations are defined as $u_i' = u_i - \langle u_i \rangle$, where $u_i$ denotes the instantaneous velocity component and $\langle \cdot \rangle$ represents ensemble averaging in time. Subscripts $i\in\{z,r,\theta\}$ correspond to the streamwise, radial, and out-of-plane velocity components, respectively. Stereo-PIV data are used to evaluate mean flow quantities and turbulence statistics, whereas the high-speed planar PIV data quantify temporal dynamics, spatial coherence, and transport of separated-flow structures. Spatial averaging is denoted by an overbar, $\overline{(\cdot)}$, corresponding to averaging over the specified spatial line or region of interest (ROI).

Statistical convergence of the measured quantities is verified by evaluating the convergence of mean values and fluctuations as the ensemble size increases. For the stereo-PIV measurements, the ensemble size of 3000 realizations ensures convergence of the mean velocity components and Reynolds stresses throughout the domain. The high-speed planar PIV data do not provide sufficient statistically independent realizations for converged mean quantities. However, they are used for temporal and space--time analyses that do not depend on mean convergence. Axisymmetry of the mean flow is assessed by comparing velocity and turbulence statistics on either side of the pipe centerline, and shows excellent agreement. Symmetry in the out-of-plane direction is further supported by the consistency of out-of-plane velocity statistics. The upstream flow is verified to be fully developed prior to the expansion by comparing mean velocity profiles and turbulence intensities ($\sim5\%$) with canonical turbulent pipe-flow distributions at the corresponding Reynolds numbers. The incoming boundary layer at the expansion plane is fully turbulent, with no large-scale distortions that influence the separation process. A detailed discussion of statistical convergence, symmetry verification, and inlet-flow characterization under the current experimental conditions is provided in \citet{jose2026effect}. The uncertainty in mean velocity components is less than 0.5\% of $U_m$ and less than 0.2\% of $U_m^2$ for Reynolds stress terms from the stereo-PIV data. The corresponding uncertainty in instantaneous velocity from the time-resolved measurements is approximately 0.8\% of $U_m$. Additional details on uncertainty estimation are provided in Appendix~\ref{app:uncertainty}.


\section{Results}

\subsection{Mean and instantaneous flow topology}
\label{ssec:mean_flow}

\begin{figure*}
    \centering
    \includegraphics[width=\linewidth]{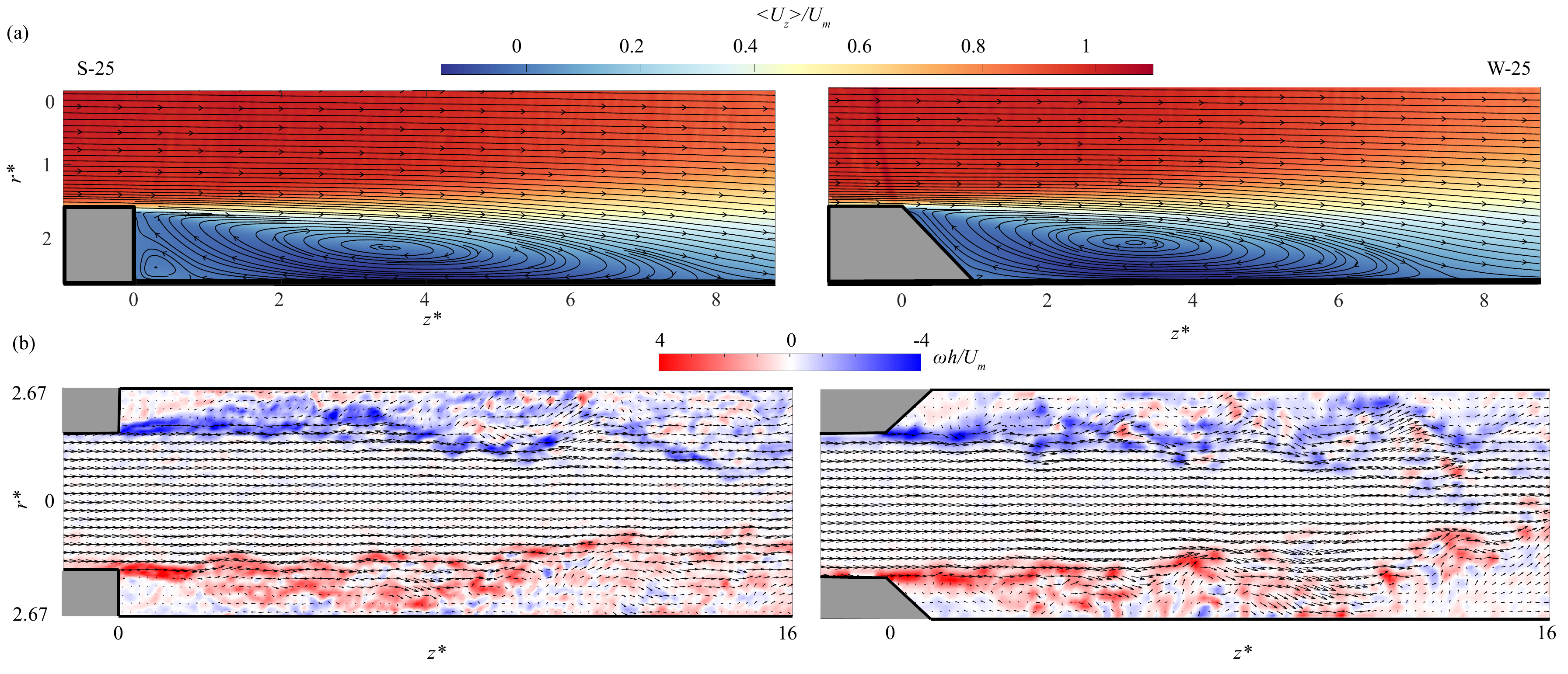}
    \caption{(a) Maps of normalized streamwise velocity for bottom half of the axisymmetric expansion for S-25 and W-25 overlaid with streamlines. (b) Normalized instantaneous vorticity, $\omega h/U_m$ for S-25 and W-25 cases across the whole axisymmetric expansion overlaid with vectors. Vectors are diluted by 3 in both directions.}
    \label{fig:Vel_vort}
\end{figure*}

Figure~\ref{fig:Vel_vort} presents the mean streamwise velocity fields together with representative sample instantaneous vorticity snapshots for the S-25 and W-25 cases. The ensemble-averaged velocity fields in Fig.~\ref{fig:Vel_vort}(a) are overlaid with streamlines, while the instantaneous snapshots in Fig.~\ref{fig:Vel_vort}(b) show velocity vectors, diluted by a factor of three in both directions. In all cases, the expansion induces flow separation and the formation of a primary recirculation region bounded by a developing shear layer. The overall topology of the separated flow is similar across geometries and Reynolds numbers in the mean fields shown in Fig.~\ref{fig:Vel_vort}(a). In particular, the mean reattachment location occurs at approximately $z^* \approx 8$ for all cases, indicating that the recovery length of the time-averaged flow is not strongly influenced by expansion slope or Reynolds number within the parameter range examined here. Nevertheless, clear geometric effects emerge in the near-corner region and in the instantaneous organization of the flow. In the step configurations, a pronounced secondary recirculation vortex forms immediately downstream of the vertical expansion face. This secondary vortex deflects the return flow away from the wall, promoting a more abrupt interaction between the recirculating fluid and the incoming shear layer. In contrast, the wedge configurations exhibit a substantially weaker and smaller secondary recirculation, and the return flow remains largely aligned with the sloped wall as it approaches the expansion corner.

The instantaneous snapshots in Fig.~\ref{fig:Vel_vort}(b) show that, despite comparable mean topologies, the two geometries produce markedly different near-corner flow organization. However, mean fields and instantaneous snapshots alone cannot determine whether these differences reflect a fundamental reorganization of coherent transport or simply a redistribution of turbulence intensity within a common dynamical framework. To address this, we examine how expansion geometry influences the spatial extent, temporal persistence, and downstream transport of unsteady motions. The following sections quantify these using turbulence statistics, spatial and temporal coherence measures, space-time correlations, and Lagrangian diagnostics, establishing how expansion slope reorganizes coherent transport in axisymmetric separated flows.

\subsection{Turbulence production near the expansion}
\label{ssec:TKE_re}

The development of turbulence downstream of the expansion is governed by the formation and growth of the separated shear layer. Previous measurements of the turbulent kinetic energy (TKE) field demonstrated consistently higher normalized TKE levels in the wedge cases compared with the step cases, together with a more rapid radial broadening of the turbulent region downstream of separation \citep{jose2026effect}. These trends indicate enhanced entrainment and a more spatially distributed turbulent field in the gradual expansion. The present analysis focuses on the near-expansion region to identify the origin of these differences.

\begin{figure*}
    \centering
    \includegraphics[width=0.98\linewidth]{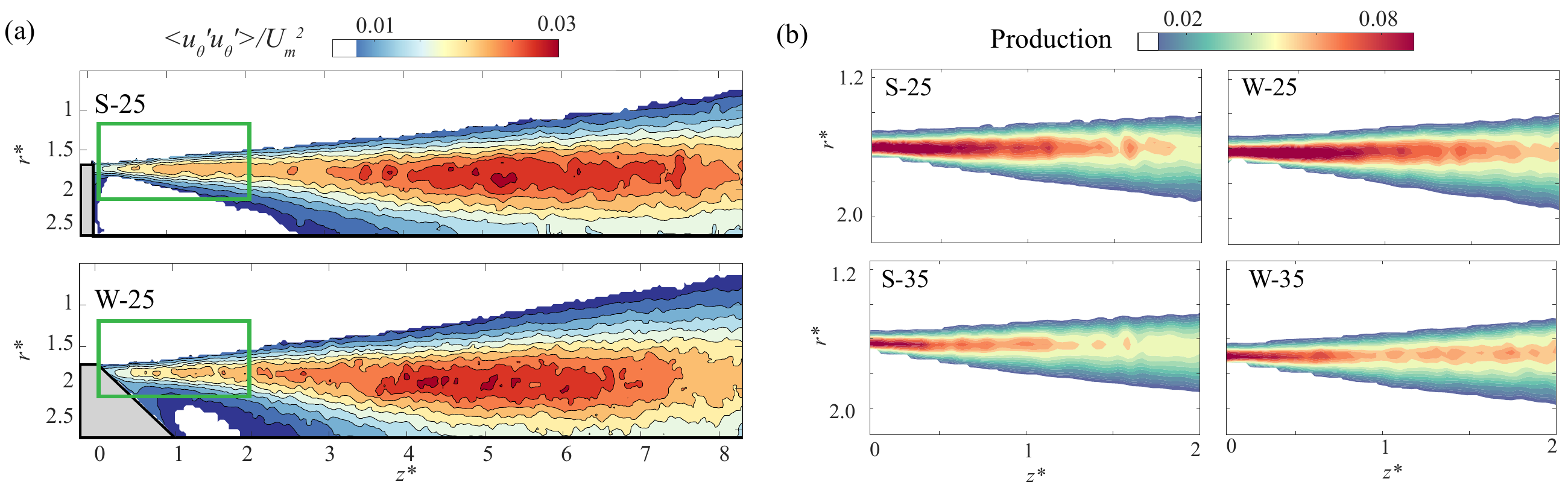}
     \caption{(a) Contour plots of normalized out-of-plane Reynolds stress term, $\langle u_\theta' u_\theta' \rangle $ for S-25, and W-25 cases. (b) Magnified map of TKE production in the immediate vicinity of the expansion at $0<z^*<2$ and $1.17<r^*<2.17$ for all four cases. Values lower than the lower limits of the colorbar are set to white in (a) and (b) for easier interpretation.}
    \label{fig:TKE_ww_prod}
\end{figure*}

Figure~\ref{fig:TKE_ww_prod}(a) shows contour maps of the normalized out-of-plane Reynolds stress, $\langle u_\theta' u_\theta' \rangle/U_m^2$, which quantifies the strength of out-of-plane velocity fluctuations. Elevated values appear immediately downstream of the expansion corner, aligned with the developing shear layer. In the wedge cases, this high-fluctuation region extends farther from the wall and occupies a larger spatial extent, whereas in the step cases it remains more localized near the separation corner. This contrast indicates that three-dimensional fluctuations are introduced earlier and over a broader region in the gradual expansion, consistent with the previously observed higher TKE levels.

Further insight is obtained from the turbulent kinetic energy production term,
\begin{equation}
\mathcal{P}_k = - \left\langle u_i' u_j' \right\rangle \frac{\partial U_i}{\partial x_j},
\label{eq:Pk}
\end{equation}
which represents the rate of energy transfer from the mean flow to turbulent fluctuations. The magnified maps in Fig.~\ref{fig:TKE_ww_prod}(b) show that production is concentrated along the separated shear layer in all cases, where strong mean velocity gradients are present. However, the wedge cases exhibit a wider, more spatially distributed production region extending farther from the wall, whereas the step cases display a thinner production band confined near the separation point. The lower-Reynolds-number cases exhibit higher normalized production levels for both geometries, consistent with the TKE trends reported previously \citep{jose2026effect}.

These results confirm that expansion geometry determines how turbulent energy is initially introduced into the flow, with the wedge producing higher TKE over a broader shear-layer region and the step confining a thinner, more intense production band near the separation corner \citep{jose2026effect}. These near-separation differences establish the initial conditions for the downstream organization of coherent structures and transport processes examined in the following sections.


\subsection{Spatial Coherence}
\label{ssec:spatial_spectra}

\begin{figure}[!]
    \centering
    \includegraphics[width=0.93\linewidth]{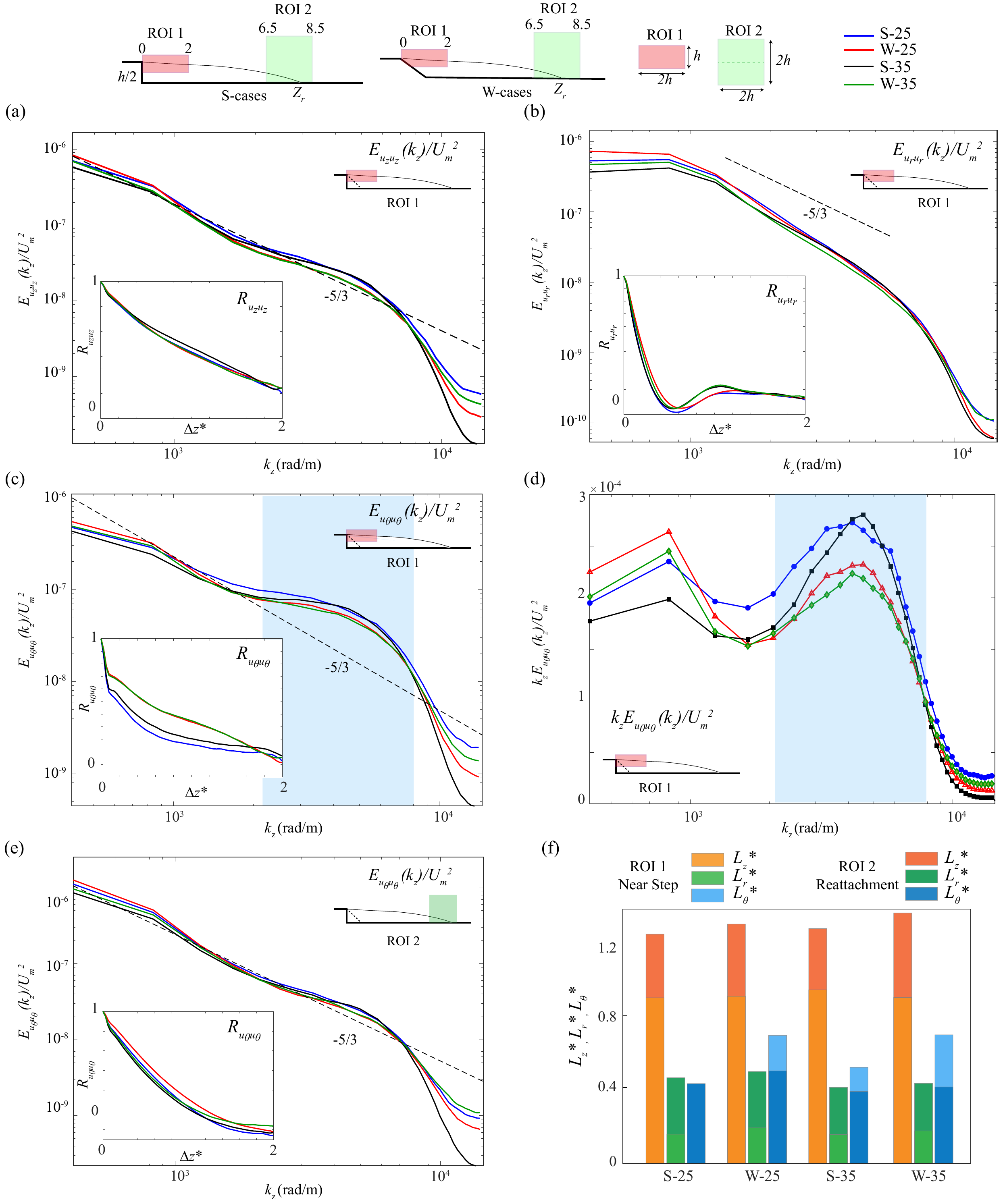}
    \caption{Normalized one-dimensional streamwise energy spectra, (a) $E_{u_zu_z}/U_m^2$, (b) $E_{u_ru_r}/U_m^2$, and (c) $E_{u_\theta u_\theta}/U_m^2$, evaluated near the expansion corner (ROI~1) with the corresponding streamwise spatial autocorrelations, $R_{u_zu_z}$, $R_{u_ru_r}$, and $R_{u_\theta u_\theta}$ in inset. (d) Premultiplied spectrum of the out-of-plane component, $k_z E_{u_\theta u_\theta}/U_m^2$, in ROI~1. (e) $E_{u_\theta u_\theta}/U_m^2$ and corresponding $R_{u_\theta u_\theta}$ (inset) in the near-reattachment region (ROI~2). (f) Normalized integral length scales, $L_z^*$, $L_r^*$ and $L_\theta^*$ in both ROIs.}
    \label{fig:energyspectra}
\end{figure}

Figure \ref{fig:energyspectra} compares the spatial organization of the velocity fluctuations between all four cases, quantified using one-dimensional streamwise autocorrelation functions and the corresponding streamwise energy spectra, evaluated along the streamwise direction $z$ within confined regions of interest (ROIs) at fixed radial locations $r$ calculated from high-resolution stereo PIV. In separated shear flows, streamwise coherence provides a direct measure of the spatial persistence and organization of energy-containing structures in the shear layer. The use of autocorrelation functions and one-dimensional spectra to characterize coherence and scale distributions in turbulent flows follows the classical approach to turbulent structure and spectral analysis \citep{tennekes1972first, Pope_2000}. In the present study, the ROIs are restricted to dynamically relevant regions, (i) near the separated shear layer (ROI~1) and (ii) near mean reattachment (ROI~2), such that the dominant flow topology does not vary appreciably within each region. ROI~2 spans a larger radial extent to account for the broadened and outwardly displaced shear layer in the vicinity of reattachment, where energetic motions occupy a wider portion of the flow. This ensures that the extracted statistics reflect the organization of fluctuations within a single flow regime, rather than combining contributions from dynamically distinct regions. Note that the dimensions of ROI~1 and ROI~2 are chosen arbitrarily but remain consistent across all cases. 

The normalized streamwise spatial autocorrelation of each velocity component is calculated from
\begin{equation}
R_{ii}(\Delta z)= 
\overline{
\frac{ \left\langle u_i'(r,z,t)\,u_i'(r,z+\Delta z,t) \right\rangle }
{ \left\langle u_i'^2(r,z,t) \right\rangle }
},
\label{eq:Rii_space}
\end{equation}
where $\Delta z$ denotes the streamwise separation, $\langle \cdot \rangle$ represents ensemble averaging, and the overbar $\overline{(\cdot)}$ denotes spatial averaging over the specified ROI (over radial extent). In the present analysis, the spatial autocorrelation is first evaluated locally along each streamwise line and subsequently averaged over the radial extent of the ROI. We apply spatial averaging to reduce the variance and suppress sensitivity to localized intermittency and small-scale inhomogeneity, thereby capturing the dominant streamwise correlation structure. Examining individual streamwise lines within each ROI yields integral length scales and spectral PIV locations that are comparable, confirming that spatial averaging primarily improves statistical convergence without altering the underlying scale distribution or biasing the dominant coherent signature. The decay of $R_{ii}(\Delta z)$ quantifies the streamwise coherence, and the mean characteristic lengths presented in Fig.~\ref{fig:energyspectra}(f) for each ROI are obtained by integrating $R_{ii}(\Delta z)$ up to its first zero crossing. These integral length scales provide a compact measure of the streamwise extent of the dominant energy-containing motions within each region.

Complementary information is provided by the one-dimensional streamwise energy spectra, $E_{ii}(k_z)$, which describe the distribution of velocity variance as a function of streamwise wavenumber. We have used the standard definition of energy spectra calculated from
\begin{equation}
E_{ii}(k_z)= 
\overline{
\left\langle \left| \hat{u}_i(k_z,t) \right|^2 \right\rangle
},
\qquad 
\hat{u}_i(k_z,t)= 
\int u_i'(r,z,t)\, \mathrm{e}^{-\mathrm{i} k_z z}\,\mathrm{d}z ,
\label{eq:Eii}
\end{equation}
where $k_z$ is the streamwise wavenumber and $\hat{u}_i$ denotes the spatial Fourier transform of $u_i'$ along $z$. As with the correlations, spectra are first computed locally and then spatially averaged within the ROI to obtain representative scale distributions for the separated shear layer and recirculation region. Figure~\ref{fig:energyspectra} presents the streamwise spectra and spatial coherence of the velocity fluctuations in the shear layer for all 4 cases. One-dimensional streamwise energy spectra of the streamwise, radial, and out-of-plane velocity components are evaluated near the expansion corner (ROI~1) in Fig.~\ref{fig:energyspectra}(a)-(c), together with the corresponding streamwise spatial autocorrelations as insets. The premultiplied spectra of Fig.~\ref{fig:energyspectra}(c), $k_z E_{u_\theta u_\theta}(k_z)/U_m^2$, is presented in Fig.~\ref{fig:energyspectra}(d).  The out-of-plane energy spectrum and corresponding spatial autocorrelation in the near-reattachment region (ROI~2) are shown in Fig.~\ref{fig:energyspectra}(e). Together, the autocorrelations and energy spectra provide a statistically robust and physically consistent framework for quantifying how expansion geometry modulates the size and coherence of energetic motions in the separated flow.

The spatial spectra and autocorrelations in Fig.~\ref{fig:energyspectra} quantify the streamwise organization of the separated shear layer across velocity components. In ROI~1, the streamwise and radial spectra (Figs.~\ref{fig:energyspectra}(a) and (b)) exhibit broadly similar intermediate-wavenumber behavior across geometries and Reynolds numbers, without the emergence of distinct geometry-specific spectral peaks. The streamwise component displays an approximate $-5/3$ decay over an intermediate range of wavenumbers, whereas the radial spectra do not follow a clear inertial-range scaling. This distinction reflects the anisotropic and inhomogeneous nature of the separated flow, in which energy transfer and spatial organization differ between components.

A distinct feature appears in the out-of-plane spectrum $E_{u_\theta u_\theta}$ in ROI~1 (Fig.~\ref{fig:energyspectra}(c)), where a pronounced spectral hump is observed over $k_z \approx 1000$--$8000~\mathrm{rad/m}$, with a peak near $k_z \approx 5000~\mathrm{rad/m}$, corresponding to wavelengths $\lambda_z/h \approx 0.1$--$0.8$ and peaking near $\lambda_z/h \approx 0.17$. The premultiplied spectrum in Fig.~\ref{fig:energyspectra}(d) confirms that this band carries the dominant intermediate-scale energy of the developing shear layer. The hump is confined to the out-of-plane component, absent in the radial spectra (Fig.~\ref{fig:energyspectra}(b)) and only weakly present in the streamwise component (Fig.~\ref{fig:energyspectra}(a)), consistent with the preferential accumulation of out-of-plane fluctuation energy near the expansion corner evidenced by the elevated $\langle u'_\theta u'_\theta \rangle$ in Fig.~\ref{fig:TKE_ww_prod}(a). Its physical origin lies in the impingement of the return flow on the separating shear layer at the expansion corner, which promotes out-of-plane redistribution of near-separation vorticity; the streamwise and radial components, aligned with the shear-layer direction, are not subject to this mechanism and do not carry the hump. The geometry-invariant position of the hump is consistent with the
comparable shear-layer thickness at detachment across geometries \citep{jose2026effect}: because this thickness sets the preferred streamwise wavelength at separation, the hump wavenumber is established locally near the expansion plane and is insensitive to the downstream differences in shear-layer spreading rate between the step and wedge configurations. The hump appears consistently in individual streamwise-line spectra before spatial averaging, confirming it is an intrinsic structural scale of the shear layer rather than an averaging artifact. To our knowledge, this feature has not been reported previously, and it identifies a preferred band of streamwise wavenumbers at which out-of-plane energy redistribution is most active.

Across all configurations, the position of the hump remains nearly unchanged, indicating that the dominant streamwise organization scale is set by the separated shear layer itself rather than by the expansion slope. This invariance suggests that the characteristic streamwise wavelength is established near separation by the local shear-layer thickness at detachment, which remains comparable between geometries despite differences in downstream spreading. The primary geometry-dependent differences, therefore, lie in the amplitude and coherence of this band. The streamwise autocorrelations of the out-of-plane component, $R_{u_\theta u_\theta}$, shown in the insets of Figs.~\ref{fig:energyspectra}(c) and (e), quantify the spatial reach of the out-of-plane coherence. In ROI~1, the wedge cases maintain positive correlation over larger streamwise separations before reaching the first zero crossing, yielding larger integral length scales $L^*_\theta$ consistent with Fig.~\ref{fig:energyspectra}(f), while the step cases decorrelate over shorter distances despite exhibiting a sharper and more energetic spectral hump. This contrast reflects two distinct coherence architectures: in the step cases, out-of-plane energy is concentrated into a narrow wavenumber band while its spatial reach remains confined, whereas the wedge cases distribute energy more broadly across wavenumbers while sustaining coherence over larger spatial extents. Together with the larger streamwise integral length scales $L^*_z$ in the step cases (Fig.~\ref{fig:energyspectra}(f)), this reflects structures that are elongated in the streamwise direction but narrow in the cross-stream direction, consistent with the radially confined production band identified in Section~\ref{ssec:TKE_re}. The wedge, by contrast, distributes production over a broader shear-layer region, yielding a wider cross-stream coherence footprint and smaller $L^*_z$. The step cases exhibit a sharper and more energetic hump, consistent with the more intense, slowly expanding shear layer that concentrates fluctuation energy into a narrower spatial band \citep{jose2026effect}; the wedge cases display a broader, less sharply defined peak, consistent with a more distributed shear-layer energy input that spreads the out-of-plane organization over a wider range of spatial scales. Near reattachment, the $R_{u_\theta u_\theta}$ curves converge across all cases as the shear layer thickens and mean shear weakens downstream, indicating that the geometry-dependent differences in out-of-plane coherence organization diminish progressively toward reattachment.

Farther downstream, near reattachment, the out-of-plane spectra presented in Fig.~\ref{fig:energyspectra}(e) show that the prominence of the hump decreases and the spectra collapse more closely across cases. The corresponding integral length scales (Fig.~\ref{fig:energyspectra}(f)) converge accordingly, indicating that the influence of expansion geometry diminishes as the shear layer thickens and reorganizes toward reattachment. The persistence of the hump at approximately the same $k_z$, albeit with reduced amplitude, suggests that the intermediate streamwise organization scale is intrinsic to the separated shear layer, while its strength and spatial coherence are modulated by the expansion geometry. Overall, the spatial analysis presented in Fig.~\ref{fig:energyspectra} demonstrates that expansion slope reorganizes the coherence and localization of shear-layer structures without introducing new characteristic streamwise scales.

\subsection{Temporal coherence}
\label{ssec:temporal_coherence}

\begin{figure*}
    \centering
    \includegraphics[width=0.95\linewidth]{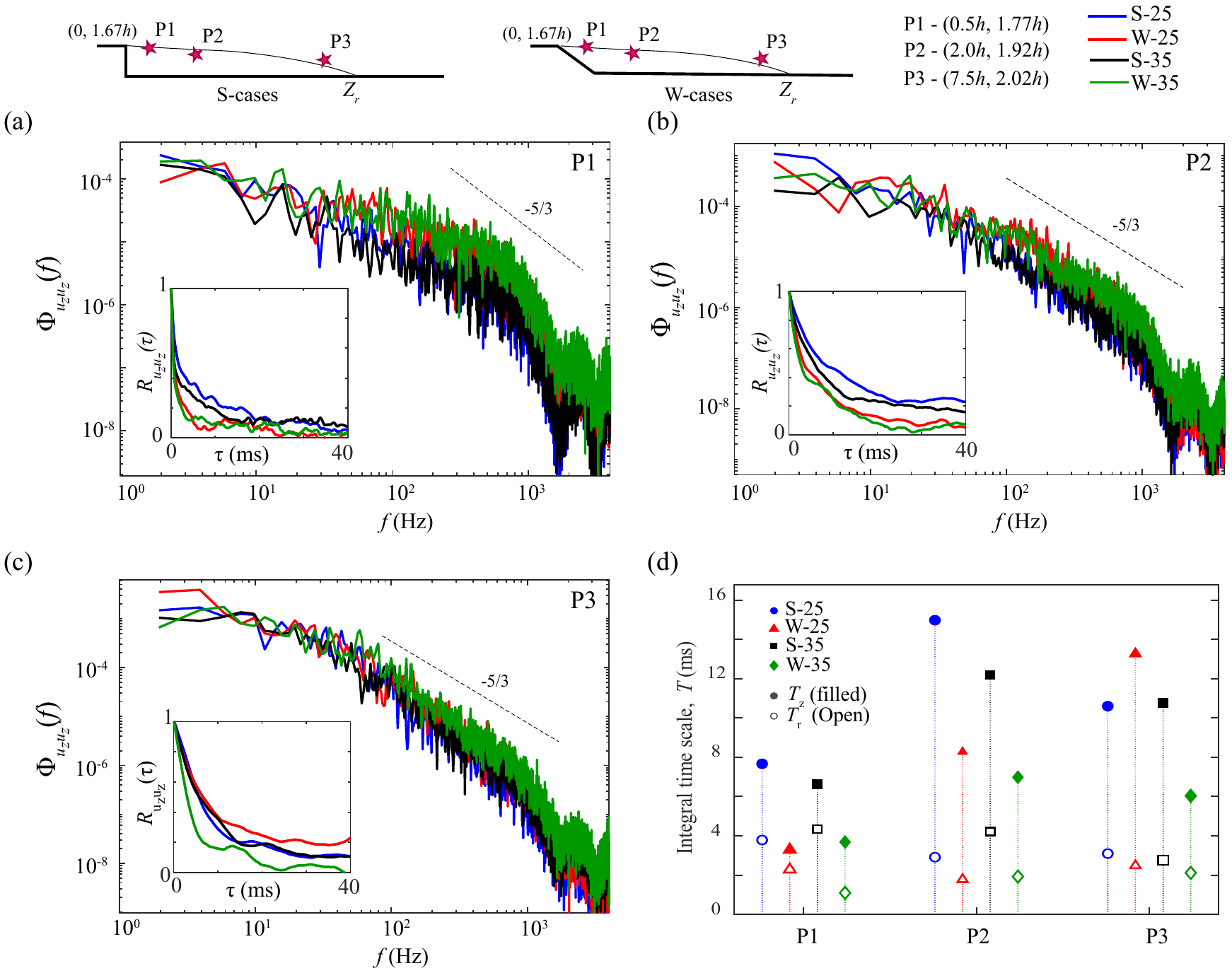}
    \caption{Normalized temporal energy spectra, $\Phi_{u_zu_z}/U_m^2$, evaluated in (a) P1 (0.5$h$,1.77$h$), (b) P2 (2.0$h$,1.92$h$) and (c) P3 (7.5$h$,2.02$h$) plotted as a function of frequency $f$. Insets show the corresponding temporal autocorrelations, $R_{u_zu_z}(\tau)$. The schematic above defines the locations of the P1, P2 and P3 relative to the flow field. (d) The integral time scales, the streamwise time scale $T_z$ (filled markers) and the radial time scale $T_r$ (open markers), for all 3 points.}
    \label{fig:Temp_spectra}
\end{figure*}

To analyze the temporal dynamics of the separated shear layer, we examine the temporal autocorrelation functions, integral time scales, and temporal energy spectra of the velocity fluctuations calculated from high-speed planar PIV data.
The normalized temporal autocorrelation is calculated as
\begin{equation}
R_{u_i u_i}(\tau)=
\frac{
\left\langle u_i'(t)\,u_i'(t+\tau) \right\rangle
}{
\left\langle u_i'^2 \right\rangle
},
\label{eq:Rii_time}
\end{equation}
where $\tau$ denotes the time delay. The integral time scale is obtained by integrating $R_{u_i u_i}(\tau)$ up to its first zero crossing. The temporal energy spectrum is given by
\begin{equation}
\Phi_{ii}(f)=
\left\langle \left| \hat{u}_i(f) \right|^2 \right\rangle , 
\qquad
\hat{u}_i(f)=
\int u_i'(t)\,\mathrm{e}^{-2\pi \mathrm{i} f t}\,\mathrm{d}t ,
\label{eq:Phi_ii}
\end{equation}
where $f$ denotes frequency.
The temporal organization of the separated shear layer provides a complementary measure of the dynamics we identified above from the spatial analysis. Figure~\ref{fig:Temp_spectra} presents the temporal energy spectra $\Phi_{u_z u_z}(f)$, the corresponding temporal autocorrelations $R_{u_z u_z}(\tau)$, and the integral time scales at 3 key points P1-P3 shown in the top panel of the figure. These points are chosen along the ensemble-averaged shear-layer centerline, with P1 near the separation, P2 further downstream, and P3 near the reattachment point. At all three locations (P1-P3), the temporal spectra of the streamwise velocity fluctuations, $\Phi_{u_z u_z}(f)$, remain broadband and do not exhibit geometry-specific spectral peaks. The absence of distinct spectral peaks indicates that the expansion slope does not introduce new characteristic temporal frequencies. Figure \ref{fig:Temp_spectra}(a) shows that closer to the separation, the spectra remain dominated by large-scale energy and do not collapse onto the expected $-5/3$ scaling. However, as the coherent vortices begin to break down, further downstream, the decay in the intermediate-frequency range follows an approximate $-5/3$ power law.

Further examining the results at P1 (Fig.~\ref{fig:Temp_spectra}(a)) shows that the temporal spectra of the step and wedge cases are closely aligned at low frequencies, indicating comparable large-scale temporal energy content immediately downstream of separation. Differences become more apparent at intermediate and higher frequencies, where the wedge cases exhibit elevated spectral levels relative to the step cases, while the overall broadband character and inertial-range scaling remain similar across configurations. The temporal autocorrelations shown in the inset of fig.~\ref{fig:Temp_spectra}(a) indicates that the step cases maintain higher correlation levels at intermediate time delays compared with the wedge cases, consistent with the redistribution of fluctuation energy toward intermediate frequencies seen in the spectra. Examining the integral time scale at P1  shown in Fig.~\ref{fig:Temp_spectra}(d), reflecting that the influence of expansion slope on temporal persistence is weak immediately downstream of separation and becomes more pronounced farther into the developing shear layer. Taken together, these trends suggest that expansion geometry influences both the distribution of fluctuation energy across frequencies and the temporal persistence of coherent motions, without establishing new dominant temporal scales.

The geometry-dependent differences are more clearly expressed downstream at P2 ($z^*=2$, Fig.~\ref{fig:Temp_spectra}(b)). Here, the temporal autocorrelations in the inset show slower decorrelation in the step cases, and the integral time scales in Fig.~\ref{fig:Temp_spectra}(d) are correspondingly larger than those of the wedge cases, indicating that streamwise fluctuations persist longer in the step geometry at this downstream location. These longer integral time scales alongside the narrower spatial coherence of the step cases are consistent with the trends identified in Section~\ref{ssec:spatial_spectra}. In the step cases, a thin and intense shear layer confines energetic fluctuations to a narrow radial band, so a fixed measurement point within that band continuously samples a concentrated and persistent fluctuation signature, producing elevated temporal correlation. In the wedge cases, the broader production region distributes fluctuation energy over a wider radial extent, reducing the temporal persistence sampled at any fixed point. Inertial-range spectral levels also remain consistently higher in the step cases at this location, in addition to the spectra remaining elevated across the inertial range relative to the wedge cases.

Farther downstream, near reattachment, P3 (Fig.~\ref{fig:Temp_spectra}(c)), the spectra and temporal autocorrelations collapse more closely across configurations, and the differences between the step and wedge cases are substantially reduced relative to locations P1 and P2. Similarly, the autocorrelation functions also exhibit similar trends and more uniform integral time scales (see Fig.~\ref{fig:Temp_spectra}(d)).

The Reynolds-number dependence of the integral time scales in Fig.~\ref{fig:Temp_spectra}(d) shows that both $T_z$ and $T_r$ are consistently larger at $Re_h = 25{,}000$ than at $Re_h = 35{,}000$ at all measurement locations, reflecting the longer convective time scales associated with the lower inlet velocity at fixed step height. Despite this shift in magnitude, the relative ordering between the step and wedge configurations remains unchanged across Reynolds numbers, indicating that the influence of expansion slope on temporal persistence is largely independent of the Reynolds-number range examined here. Across all cases and locations, the radial time scale $T_r$ remains systematically smaller than the streamwise time scale $T_z$, reflecting the shorter decorrelation time of radial fluctuations, which are more strongly associated with the intermittent passage of vortical structures across the measurement point than with the sustained streamwise-aligned coherence that governs $T_z$.

\subsection{Space--time coherence}
\label{ssec:spacetime_Uc}

\begin{figure*}
    \centering
    \includegraphics[width=0.95\linewidth]{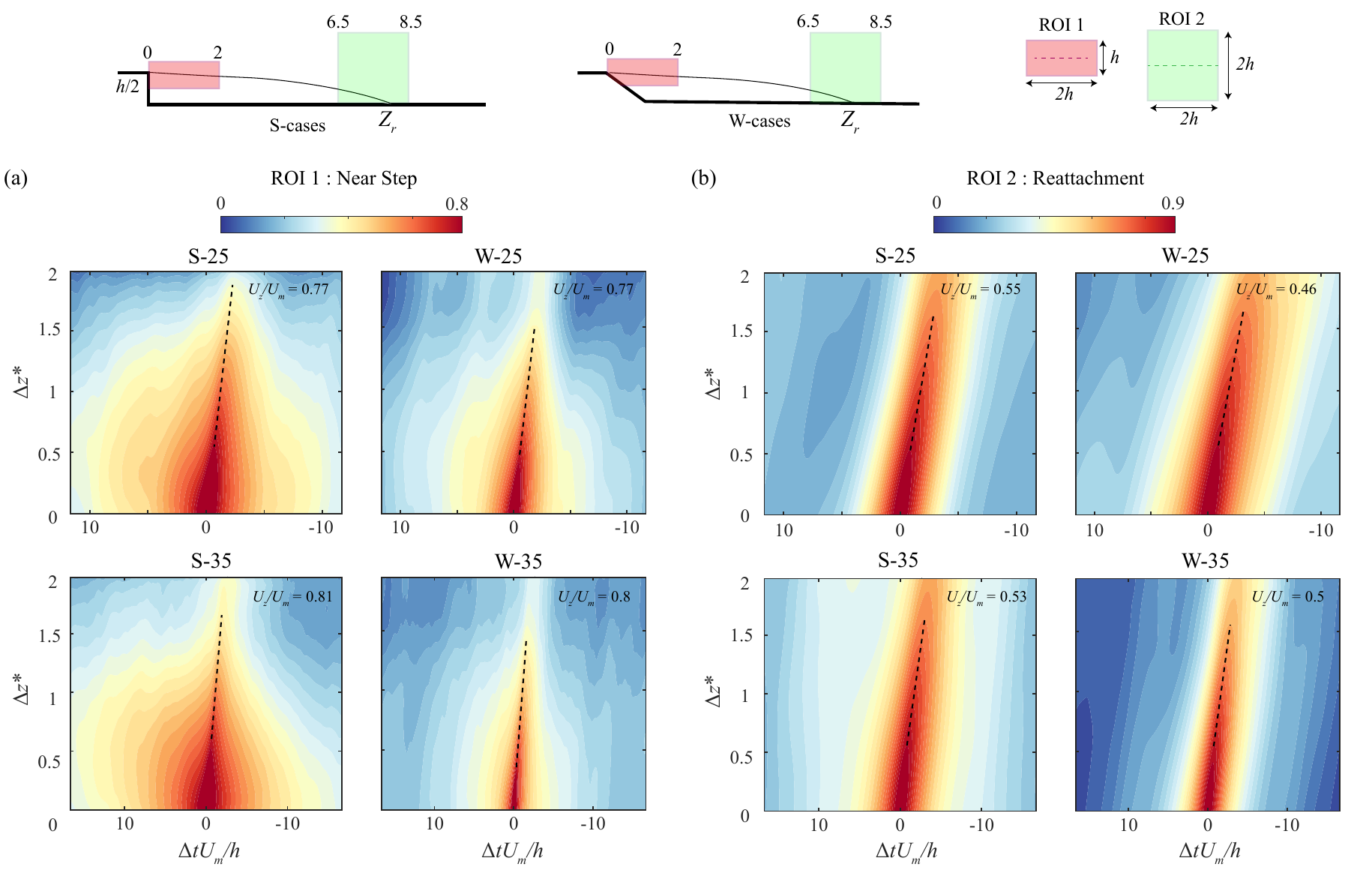}
    \caption{Space-time correlations of the streamwise velocity fluctuations in the separated flow. (a),(b) Normalized space-time correlation maps, $R_{u_zu_z}(\Delta z,\Delta t)$, evaluated in the near-expansion region (ROI~1) and the near-reattachment region (ROI~2), respectively, for all four cases. The dashed lines highlight the dominant convection paths. A schematic of the chosen ROIs is provided above and matches those defined in Fig.~\ref{fig:energyspectra}.}
    \label{fig:space_time}
\end{figure*}

The downstream transport of coherent structures is examined using the space-time correlation of the streamwise velocity fluctuations, shown in Fig.~\ref{fig:space_time}. While the preceding spatial and temporal analyses quantified the size and persistence of coherent motions separately, they do not directly determine how these motions propagate downstream. The space-time formulation resolves the coupled evolution in space and time and therefore enables direct estimation of the convection velocity of the dominant shear-layer structures. The correlation is calculated as
\begin{equation}
R_{u_z u_z}(\Delta z,\Delta t)=
\overline{
\frac{
\left\langle
u_z'(r,z,t)\,u_z'(r,z+\Delta z,t+\Delta t)
\right\rangle
}{
\left\langle u_z'^2(r,z,t) \right\rangle
}
},
\label{eq:Rii_spacetime}
\end{equation}
and is evaluated within the same ROIs as in section \ref{ssec:spatial_spectra}. As in the spatial analysis, the correlation is first computed locally along each streamwise line and subsequently averaged over the radial extent of the ROI in order to reduce sensitivity to localized intermittency while preserving the dominant convection signature of the shear-layer fluctuations. The resulting $(\Delta z,\Delta t)$ correlation maps in Fig.~\ref{fig:space_time} reveal coherent motions as inclined ridges of elevated correlation. The ridge inclination provides an estimate of the convection velocity associated with the dominant streamwise fluctuation signatures, while the ridge width reflects the range of convection velocities associated with the advecting shear-layer structures. When expressed in normalized coordinates, $\Delta z^*=\Delta z/h$ and $\Delta t\,U_m/h$, the ridge slope yields the normalized convection velocity $U_c/U_m$, where $U_c$ represents a characteristic convection velocity within each ROI obtained from the dominant ridge location after radial averaging.

Immediately downstream of separation, in ROI~1 (Fig.~\ref{fig:space_time}(a)), all cases exhibit a well-defined inclined ridge with similar inclination, yielding $U_c/U_m \approx 0.77$--$0.81$. These values are higher than convection velocities typically reported within the reattaching shear layer of planar backward-facing step configurations (0.4–0.6$U_m$), reflecting that in ROI 1 located in the near-expansion region, the shear layer has not yet expanded significantly into the low-momentum recirculation zone. This can be attributed to the fact that close to the separation point, at $z^*$ = 0–2, the mean streamwise velocity above the step height is still very close to $U_m$, and the dominant fluctuation signatures are convected primarily by this high-momentum outer flow. Downstream, in ROI 2 (Fig.~\ref{fig:space_time}(b)), the normalized convection velocities reduce to $U_c/U_m \approx$ 0.46–0.55, and better agree with prior studies \citep{furuichi2003spatial, troutt1984organized}. Although the spatial and temporal analyses revealed geometry-dependent differences in coherence strength and persistence near separation, the space-time maps in Fig.~\ref{fig:space_time} indicate that the characteristic downstream transport speed of the coherent structures remains similar for both ROIs. 

Although convection velocities are similar across all cases in both ROIs, the space--time maps reveal a notable geometry-dependent difference in the width of the correlation ridge in ROI~1. The step cases exhibit a broader band of elevated correlation surrounding the dominant ridge, whereas the wedge cases display a more narrowly confined ridge structure. This difference reflects the distinct momentum states of the return flow entering the shear layer from below. In the step geometry, the return flow loses substantial momentum before reaching the shear layer through two compounding effects: the secondary recirculation vortex at the expansion foot actively draws momentum from the returning fluid, and the flow undergoes a $90^\circ$ directional change before encountering the free-stream. In the wedge geometry, neither mechanism operates with comparable strength; the return flow follows the sloped wall closely, undergoes a smaller directional change, and arrives at the shear layer with significantly higher momentum. The radially averaged space--time correlation in ROI~1 therefore captures contributions from both the high-momentum outer flow and the momentum-depleted fluid parcels entrained from below, producing a wide spread of local convection velocities that broadens the ridge in the step cases. The wedge cases, with a more uniformly energetic return flow, produce a narrower spread of local convection velocities and a correspondingly more confined ridge structure. This ridge broadening is therefore a geometric signature of the return-flow dynamics rather than a reflection of reduced transport coherence, and it diminishes downstream as the influence of the expansion corner weakens, consistent with the convergence observed in the spatial and temporal coherence measures.

\subsection{Lagrangian coherent structures}
\label{ssec:lcs}

\begin{figure*}
    \centering
    \includegraphics[width=\linewidth]{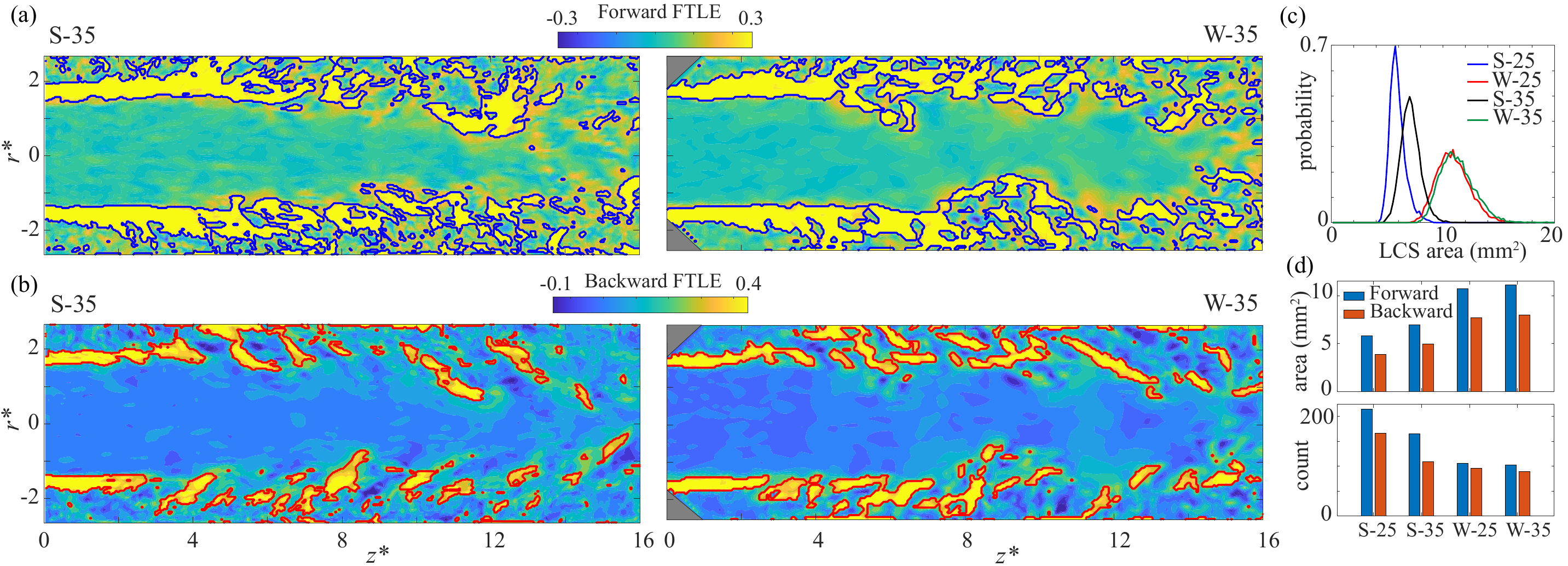}
    \caption{(a) Sample forward FTLE overlaid with repelling LCS (blue) and (b) backward FTLE with attracting LCS (red) for S-35 and W-35 cases. The data is super-sampled by a factor of two in $r-z$ directions, the time of integration is five instantaneous snapshots and the threshold for LCS ridge detection is 60\% of max FTLE. (c) Probability distribution of the sizes of LCS structures and (d) mean size distributions and counts of LCS for all four cases.}
    \label{fig:ftle}
\end{figure*}

Finite-Time Lyapunov Exponents (FTLE) analysis is employed to examine differences in the short-time Lagrangian transport organization of the separated flow \citep{shadden2005definition,mathur2007uncovering,haller2015lagrangian}. Whereas the preceding sections characterized fluctuation organization using Eulerian measures, the FTLE framework evaluates finite-time material deformation induced by the measured unsteady velocity field. In this sense, FTLE provides a transport-oriented counterpart to the Eulerian coherence results by quantifying how nearby fluid elements separate over a finite time interval, thereby highlighting deformation patterns that shape mixing and pathway connectivity. 

FTLE is calculated for every time step for the planar high speed PIV where each time step determines new initial condition $\mathbf{x}_0=(r_0,z_0)$ at time $t_0$, the corresponding local particle trajectory $\mathbf{x}(t;t_0,\mathbf{x}_0)$ is obtained by time step integration through the consequent velocity snapshots numerically using
\begin{equation}
\mathbf{F}_{t_0}^{t_0+T}(\mathbf{x}_0) \equiv \mathbf{x}(t_0+T;t_0,\mathbf{x}_0)
= \mathbf{x}_0 + \int_{t_0}^{t_0+T}\mathbf{u}\!\left(\mathbf{x}(t;t_0,\mathbf{x}_0),t\right)\,dt ,
\end{equation}
where $T$ is the integration time window. $T$ can be chosen as a positive time to provide forward-time FTLE or as a negative time to provide backward-time FTLE. The deformation gradient $\nabla \mathbf{F}_{t_0}^{t_0+T}$ yields the Cauchy--Green tensor
\begin{equation}
\mathbf{C}_{t_0}^{t_0+T}(\mathbf{x}_0)=\left(\nabla \mathbf{F}_{t_0}^{t_0+T}\right)^{\!\top}
\left(\nabla \mathbf{F}_{t_0}^{t_0+T}\right),
\end{equation}
from which the FTLE field is computed as
\begin{equation}
\lambda_T(\mathbf{x}_0)=\frac{1}{|T|}\,
\ln \sqrt{\lambda_{\max}\!\left(\mathbf{C}_{t_0}^{t_0+T}(\mathbf{x}_0)\right)},
\end{equation}
where $\lambda_{\max}$ is the largest eigenvalue of $\mathbf{C}_{t_0}^{t_0+T}$. Forward-time FTLE emphasizes locally repelling material structures, whereas backward-time FTLE emphasizes attracting structures that delineate convergent transport pathways \citep{haller2015lagrangian}. The integration time was set to five instantaneous snapshots, corresponding to $T = 0.625\,\mathrm{ms}$, which limits the integration interval to approximately $0.37\,\tau_\eta$, where $\tau_\eta$ is the Kolmogorov time scale. At this integration time, the resulting FTLE fields faithfully represent the short-time material deformation driven by the resolved, energy-containing motions of the shear layer \cite{huang2022lagrangian}, which are used here to compare the spatial organization of instantaneous deformation patterns across expansion geometries. The resulting FTLE maps are presented in Fig.~\ref{fig:ftle}, showing the organization of the same energy-containing shear-layer structures examined in the preceding Eulerian analysis, providing a Lagrangian counterpart to those results. Because the present FTLE analysis is based on planar high-speed PIV, the reconstructed material trajectories are restricted to the $(r,z)$ plane and therefore underestimate the full three-dimensional deformation of the flow. The results are thus interpreted comparatively rather than quantitatively. Since both expansion geometries are evaluated under identical planar measurement conditions, the geometry-dependent differences in ridge area and fragmentation primarily reflect differences in the in-plane organization of material transport. The persistence of these trends across both integration directions and Reynolds numbers further supports the robustness of the comparison within the constraints of the planar formulation.

Figure~\ref{fig:ftle}(a) and (b) provide representative forward- and backward-time FTLE fields for the S-35 and W-35 cases, illustrating the characteristic transport organization at the higher Reynolds number, where geometry-dependent differences are most clearly expressed. In both geometries, the FTLE fields exhibit intermittent ridge networks concentrated within the separated shear layer and along the near-wall return flow. The ridge patterns evolve in time and remain spatially irregular, consistent with the broadband, non-periodic dynamics identified in the preceding analyses. To quantify this transport organization, contiguous regions exceeding a consistent threshold are identified as individual ridge patches. A threshold of 60\% of the maximum FTLE value within each realization is employed to isolate the dominant short-time deformation regions while excluding low-amplitude background strain. Defining the threshold relative to the instantaneous maximum ensures consistent normalization across realizations and across geometries, thereby enabling comparison of spatial connectivity independent of absolute FTLE magnitude. The area of each detected patch is computed as a measure of the spatial extent of short-time deformation regions, and statistics compiled over $10^4$ realizations for all four cases are summarized in Fig.~\ref{fig:ftle}(c) and (d).

The probability distributions in Fig.~\ref{fig:ftle}(c) show a systematic geometry dependence that is consistent across both Reynolds numbers and both integration directions. For both Reynolds numbers, the wedge cases produce significantly larger LCS patches, whereas Fig.~\ref{fig:ftle}(d) shows that they are fewer in number. The step cases exhibit smaller characteristic areas and higher patch counts. The consistency of this trend across both forward- and backward-time FTLE fields indicates that it reflects a general difference in deformation organization rather than a distinction between repelling and attracting structures. These results are consistent with the Eulerian observations and strengthen the interpretation developed above. In the wedge cases, turbulence production extends over a broader shear-layer region, and the vortical structures that develop over this wider production region manifest as larger and more contiguous deformation regions in the FTLE fields. In the step cases, production is more localized near the expansion corner and near-wall intermittency is stronger, so the corresponding FTLE fields exhibit a more fragmented pattern with shorter spatial coherence lengths. Importantly, these differences in LCS size and fragmentation arise without any change in the dominant convection velocity, reinforcing that the expansion slope primarily reorganizes the spatial extent and connectivity of deformation regions rather than altering their propagation speed.

\subsection{Near-wall transport dynamics }
\label{ssec:nearwall_convection}

\begin{figure*}
    \centering
    \includegraphics[width=0.95\linewidth]{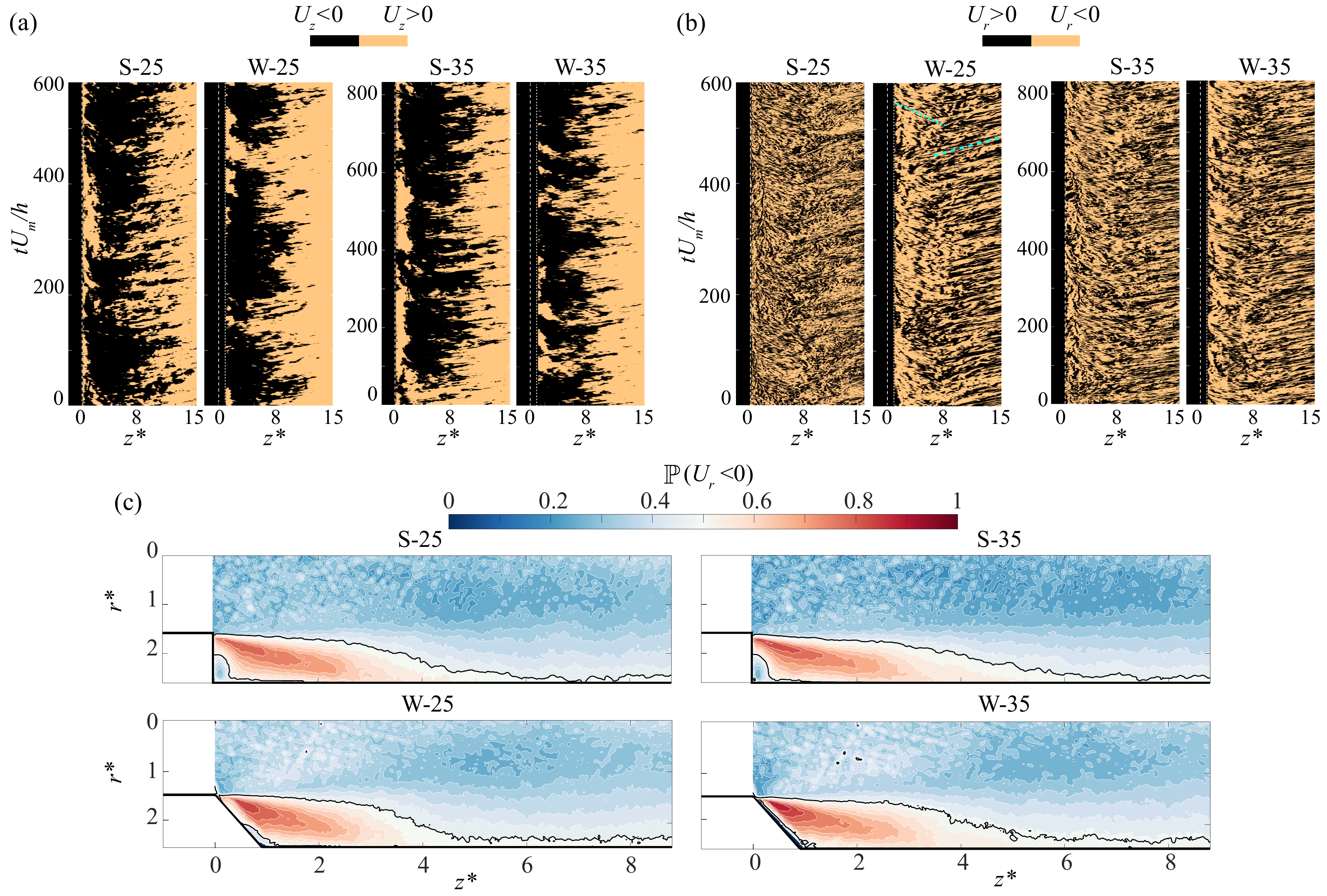}
    \caption{Binary spatio-temporal maps of (a) $U_z$ and (b) $U_r$ showing the locations of positive and negative flow directions evaluated near the wall at $r^*=2.61$ for all four cases. The dashed vertical line denotes the location of the step ($z^*=0$), and the dotted vertical line in the wedge cases corresponds to the bottom edge of the wedge at $z^*=1$. The sloped green dashed lines in (b) are illustrative guides indicating the propagation direction and apparent convection velocity of rolling near-wall structures in the $z^*$–$t$ plane. Opposite slopes correspond to downstream (positive) and upstream (reverse-flow) propagation relative to the mean flow. (c) Spatial distribution of the mean probability, $\mathbb{P}(U_r < 0)$, for the four cases obtained from the stereo-PIV data. The $50\%$ contour line is highlighted in black.}
    \label{fig:trans_prob}
\end{figure*}

The shear-layer structures identified in the preceding sections are eventually convected to the wall downstream of reattachment, where their footprint can be traced in near-wall velocity statistics. The secondary recirculation vortex present in the step geometry, which depletes return-flow momentum near the expansion foot and sustains a concentrated upstream return-flow signature at the expansion face, is expected to produce a persistent asymmetry in near-wall momentum distribution that is absent in the wedge
cases~\citep{jose2026effect}. The following diagnostics quantify how these geometry-dependent differences, established at the separation corner, manifest in the
near-wall flow after reattachment.

Figure~\ref{fig:trans_prob} presents the near-wall spatio-temporal probability as well as the radial flow direction probability maps within and downstream of the recirculation region. Whereas the preceding sections focused on shear-layer organization and bulk transport, these statistics provide an indicator of how these motions interact with the wall and how momentum is redistributed after separation. Figures~\ref{fig:trans_prob}(a) and (b) present binary spatio-temporal maps of the streamwise and radial velocity components, $U_z$ and $U_r$, extracted from the high-speed PIV measurements at a fixed near-wall location ($r^*=2.61$). The horizontal axis denotes the streamwise coordinate $z^*$, and the vertical axis represents the normalized time, $tU_m/h$, spanning $10^4$ realizations. Figure~\ref{fig:trans_prob}(c), obtained from 3000 stereo-PIV realizations, shows the spatial distribution of the mean probability $\mathbb{P}(U_r<0)$, which quantifies the prevalence of radial motion towards the centerline, indicating locations associated with significant vortical activity.

The presence of a counter-rotating secondary vortex is also highly transient and appears as intermittent regions of $U_z > 0$ in the immediate vicinity of the expansion foot. The size of this vortex fluctuates significantly, and at times it disappears altogether. Its occurrence and spatial extent are more pronounced in the step configurations, consistent with the mean flow field shown in Fig.~\ref{fig:Vel_vort}(a). In the wedge cases, the vortex can also appear intermittently; however, it remains short-lived and therefore contributes only weakly to the mean flow structure shown in Fig.~\ref{fig:Vel_vort}(a). Time maps of $U_z$ in Fig.~\ref{fig:trans_prob}(a) highlight the strongly unsteady nature of the recirculation region. While the mean flow indicates a recirculation zone extending from $z^* = 0$ to $z^* \approx 8$, the instantaneous behavior deviates substantially from this average picture. The reattachment location undergoes large excursions in time, with the separated region occasionally extending as far downstream as $z^* \approx 15$, while at other times the recirculation collapses and the flow remains fully attached over the entire measurement domain. The size of this vortex fluctuates significantly, and at times it disappears altogether. Its occurrence and spatial extent are more pronounced in the step configurations, consistent with the mean flow field shown in Fig.~\ref{fig:Vel_vort}(a). In the wedge cases, the vortex can also appear intermittently; however, it remains short-lived and therefore contributes only weakly to the mean flow structure shown in Fig.~\ref{fig:Vel_vort}(a). 

The $U_r$ spatio-temporal maps in Fig.~\ref{fig:trans_prob}(b) show inclined streaks that indicate coherent vortical motion along the wall. The slope of these streaks in the $z^*$--$tU_m/h$ plane provides an estimate of the propagation speed of rolling vortical motions, with positive slope corresponding to downstream propagation after reattachment and negative slope to upstream propagation within the recirculation. These slopes are measured by identifying the dominant ridge orientation of the inclined streak patterns, as illustrated by the green dashed lines in Fig.~\ref{fig:trans_prob}(b). The near-wall propagation speeds are slower than the bulk convection velocities reported in Section~\ref{ssec:spacetime_Uc}. In the step cases, upstream motion propagates at approximately $0.15$--$0.2\,U_m$ while downstream motion occurs at $0.35$--$0.45\,U_m$, giving an asymmetry of roughly $0.2$--$0.3\,U_m$ between the two directions. In the wedge cases, the corresponding velocities are closer in magnitude, approximately $0.2$--$0.25\,U_m$ upstream and $0.25$--$0.3\,U_m$ downstream, reflecting a more symmetric near-wall momentum environment. This asymmetry in the step cases is consistent with the persistent secondary recirculation vortex identified in the mean flow (Section~\ref{ssec:mean_flow}), which sustains a concentrated upstream return-flow signature immediately downstream of the expansion face, reinforcing the near-wall momentum deficit and producing the observed contrast between upstream and downstream propagation speeds. The substantially weaker secondary recirculation in the wedge cases reduces this bias, bringing the two propagation speeds closer in magnitude. Beyond these quantitative differences, the wedge cases exhibit thicker and more continuous streaks, whereas the step cases display thinner and more intermittent patterns, indicating larger wall-adjacent vortical structures convecting downstream of reattachment in the gradual expansion.

Figure~\ref{fig:trans_prob}(c) further highlights the spatial extent of the vortical structures transporting along the surface downstream of reattachment. The wedge cases exhibit broader regions of elevated $\mathbb{P}(U_r<0)$, while the step cases show sharper spatial gradients and more confined high-probability zones. In all configurations, the $50\%$ contour does not terminate within the measurement domain but instead asymptotes toward the wall downstream of reattachment, indicating that rolling near-wall vortices persist throughout the measured extent. The location at which this contour approaches its asymptotic near-wall position occurs slightly earlier in the step cases than in the wedge cases, consistent with the more spatially confined and intermittent wall-adjacent structures in the abrupt geometry and with the narrower spatial coherence footprint established near separation. In this post-reattachment region, the radial extent over which $\mathbb{P}(U_r < 0) > 0.5$ spans approximately $0.1$--$0.2h$, representing the time-averaged spatial envelope of near-wall vortical activity rather than the instantaneous size of individual structures, and providing a statistical measure of the characteristic radial reach of post-reattachment rolling vortices. The consistency of this scale with the deformation regions identified in the FTLE analysis confirms that the two diagnostics are sampling the same population of wall-adjacent structures. Taken together, the near-wall results extend the transport picture developed in the preceding sections: structures generated near separation are convected downstream and reorganize into rolling wall vortices after reattachment, and the geometry-dependent differences established at the expansion corner therefore remain evident at the boundary.

\section{Conclusions}
\label{sec:conclusions}

Turbulent axisymmetric expansions with abrupt ($90^\circ$) and gradual ($45^\circ$) geometries were investigated experimentally at step-height Reynolds numbers of 25000 and 35000 using high-resolution stereo-PIV and time-resolved planar high-speed PIV. The central finding is that expansion geometries producing nearly identical mean reattachment lengths can sustain fundamentally different coherent transport organization, and that this difference originates near the expansion corner and persists through reattachment.

The dominant dynamical scales of the separated shear layer are not modified by the expansion geometry. Spatial spectra reveal a geometry-invariant spectral hump in the out-of-plane velocity fluctuations near separation, whose wavenumber is set by the local shear-layer thickness at detachment rather than by expansion slope. Temporal spectra remain broadband at all downstream locations without geometry-specific dominant frequencies, and space-time correlations indicate similar normalized convection velocities in both the near-expansion and near-reattachment regions across all cases. The geometric influence, therefore, does not manifest as a shift in characteristic scales or transport speeds.

The expansion geometry primarily controls the spatial topology through which coherent fluctuations are organized and sustained. The abrupt step confines turbulence production to a thin, intense band near the separation corner, where a persistent secondary recirculation vortex depletes the momentum of the return flow before it reaches the shear layer. The resulting coherent structures are elongated in the streamwise direction but narrow in the cross-stream direction, producing stronger spectral concentration in the out-of-plane hump, longer local integral time scales, and a broader space-time correlation ridge. This ridge broadening reflects the wide spread of local convection velocities introduced by the momentum-depleted return flow rather than any enhancement of transport coherence. The gradual wedge distributes production across a broader shear-layer region, admits higher-momentum return flow, sustains larger cross-stream coherence lengths, and produces a more narrowly confined correlation ridge.

The geometry-dependent differences established near separation persist and compound as the flow develops downstream. FTLE fields show that wedge cases consistently produce larger, more contiguous deformation regions, while step cases yield a fragmented pattern of smaller ridge patches at higher counts, a distinction that holds across Reynolds numbers and both forward and backward integration directions. Near-wall spatio-temporal maps confirm that these differences extend through reattachment. Wedge cases develop thicker, more continuous near-wall coherent structure streaks and broader regions of elevated radial-inflow probability, indicative of larger wall-adjacent vortical structures convecting downstream, while step cases display thinner, more intermittent wall signatures consistent with the narrower coherence footprint established at the expansion corner.

This study shows that mean-flow similarity does not imply transport similarity, as flows with nearly identical mean structure and convection velocities can exhibit fundamentally different organization of coherent transport. When combined with the turbulence production and Reynolds-stress anisotropy results of \citet{jose2026effect}, the results show that expansion geometry governs both the production of fluctuation energy through the return flow and its downstream organization. In the gradual expansion, the return flow remains energetic and distributed, whereas in the abrupt expansion, it is depleted and redirected, leading to a more confined transport structure. Despite similar characteristic scales and convection velocities, geometry primarily reorganizes the spatial distribution and persistence of coherence, so mean-flow similarity masks a fundamental difference in transport architecture revealed only through coherence- and Lagrangian-based diagnostics. This perspective is relevant to engineering configurations in which mixing efficiency, wall heat transfer, or pressure recovery depend on the spatial organization of turbulent transport rather than on bulk turbulence levels alone.


\begin{acknowledgments}
This work was supported by the donors of ACS Petroleum Research Fund under New Directions Grant \#65901-ND9.
\end{acknowledgments}


\appendix
\section{Uncertainty estimation}
\label{app:uncertainty}

Uncertainty estimates for the stereo-PIV measurements and all derived quantities are based on established correlation-based approaches and standard uncertainty propagation procedures \citep{sciacchitano2016piv}. A detailed description of the uncertainty methodology, including correlation-peak analysis, component-wise uncertainty propagation, and treatment of inter-component correlations, is provided in the companion study and is not repeated here. The uncertainties in the reconstructed velocity components are obtained by propagating the displacement uncertainties from each camera view through the stereo reconstruction. For the present experimental configuration and an ensemble size of 3000 realizations, the maximum uncertainties in the mean velocity components are estimated to be approximately 0.4\% of $U_m$ for $U_z$, 0.25\% for $U_r$, and 0.3\% for $U_\theta$. The resulting uncertainty in the mean turbulent kinetic energy is about 0.35\% of $U_m^2$, while uncertainties in the Reynolds stress components remain below 0.15\% of $U_m^2$ across the measurement domain. For the high-speed planar PIV, the estimated uncertainty in velocity is about 1.3\% of $U_m$.

Uncertainties in derived quantities involving spatial gradients, such as the turbulence production term $\mathcal{P}_k$, are estimated using standard propagation techniques applied to products of Reynolds stresses and mean velocity gradients. Based on representative values within the shear layer, the  relative uncertainty in individual production terms is approximately 8--16\%, yielding an overall uncertainty in $\mathcal{P}_k$ of about 10\%. These uncertainty levels are consistent with previous experimental investigations of separated turbulent flows and do not influence the comparative trends or physical interpretations presented in this work.

For high speed, the hardware-limited recording length of 2.3~s with 18700 PIV snapshots yields approximately 115–230 statistically independent realizations at the integral time scales identified in Section \ref{ssec:temporal_coherence}, which is sufficient for characterizing the dominant spectral content, correlation structure, and FTLE organization reported here, but not for converged mean velocity statistics. Convection velocities extracted from the space–time correlation ridge carry an estimated uncertainty of approximately 5\% of $U_m$. This is comparable to the geometry-dependent spread in $U_c/U_m$ and therefore supports the conclusion that convection velocities are similar across geometries.


\bibliography{References}

\end{document}